\documentclass[twocolumn,showpacs,a4paper,10pt,final,nofootinbib,aip,jcp]{revtex4-1}
\usepackage{graphicx}
\usepackage{subfigure}
\usepackage{dcolumn}
\usepackage{tabularx}
\usepackage{braket}
\usepackage{bm}
\usepackage{amsmath,amssymb,amsfonts,mathrsfs}
\usepackage{wrapfig,epsfig}
\usepackage{tikz}
\usepackage{pgf}

\begin{document}

\title{   Molecular QCA embedding in microporous materials}
\author{Alberto M. Pintus}%
\email{spintus@uniss.it}%
\author{Andrea Gabrieli}%
\author{Federico G. Pazzona}
\author{Giovanni Pireddu}%
\author{Pierfranco Demontis}%
\affiliation{Dipartimento di Chimica e Farmacia, Universit\`a degli Studi di Sassari, via Vienna 2, 01700 Sassari, Italy}

\begin{abstract}
We propose a new environment for information encoding and transmission via a novel type of molecular Quantum Dot 
Cellular Automata (QCA) wire, composed of a single row of head-to-tail interacting 2-dots molecular switches.
While most of the research in the field refers to dots-bearing molecules bound on some type of surface, forming
a bidimensional array of square cells capable of performing QCA typical functions, we propose here to embed the information 
bearing elements within the channels of a microporous matrix. In this way molecules would self-assemble in a row
as a consequence of adsorption inside the 
pores of the material, forming an encased wire, 
with the crystalline environment 
giving stability and protection to the structure. DFT calculations on a diferrocenyl carborane, previously proposed and 
synthesized by Christie \textit{et al.}\cite{lentmain}, were performed both in 
vacuum and inside the channels of 
zeolite ITQ-51, indicating that information encoding and trasmission is possible within the nanoconfined 
environment. 
\end{abstract}

\pacs{ 68.43.-h }


\maketitle

\section{Introduction \label{sec:Intro}}

Quantum Dot Cellular Automata (QCA) constitute a class of computing devices with very interesting and unique features, 
first proposed and subsequently mainly investigated, by C. S. Lent and coworkers \cite{lenttougaw93,lenttougaw97,colereview}.
In the classic definition given by these authors the basic building block of a QCA, the \textit{cell}, 
is an array of four quantum dots, here generically intended as 
charge confinement sites, arranged as the corners of a square.  
Each cell hosts two charges of the same sign, which can occupy any of the four dots, and can switch between them, 
usually via quantum tunneling. 
However, due to coulombic repulsion, the only stable configurations of the cell are the ones in which the charges are 
on opposite corners of the square, as shown in Fig. 1. If there is a sufficiently 
high barrier between states 0 and 1, as defined in Fig. 1-a, it is clear that each cell is 
capable of encoding and retaining one bit of information. 
In a typical QCA, cells are arranged as to form interacting neighborhoods where the state of each cell is dependent on the state of 
neighboring ones via inter-cell coulombic interactions. 
A slightly different and better model of cell contains extra sites capable of hosting the charges, resulting in a 
symmetric \textit{null} cell state\cite{blairlent}. This is interesting in view of devising an efficient QCA clocking scheme, 
in order for the information to flow in the system only in a user defined direction\cite{lenttougaw97}. 

A linear arrangement of QCA cells like the one in fig. 1-b is called a \textit{wire}, and is capable of transmitting information 
between its ends, as its cells switch to the minimum energy state according to the state of neighbors. 
While we are going to elaborate mainly on this type of monodimensional QCA, more complex architectures are also possible where 
information is not only stored and transmitted, but also elaborated via QCA logic gates, allowing these systems to 
perform all forms of calculation, opening in principle the way to the creation of QCA computers.
The first QCA practically built were based on Si/SiO$_{2}$ interfaces or aluminium dots\cite{orlovlentrealcell}, 
with aluminium oxide tunnel junctions having 
a diameter of several tens of nanometers, and were only able to work at cryogenic temperatures. 
However ionic or zwitterionic molecular switches\cite{bianchiswitch}, showing strong charge confinement, 
could act as cells in a molecular QCA  
capable of working at room temperature, beside  
allowing for an unprecedented step in circuit miniaturization and in the minimization of energy loss 
in information transmission and elaboration.\cite{} 

\begin{figure}
  \includegraphics[width=0.44\textwidth]{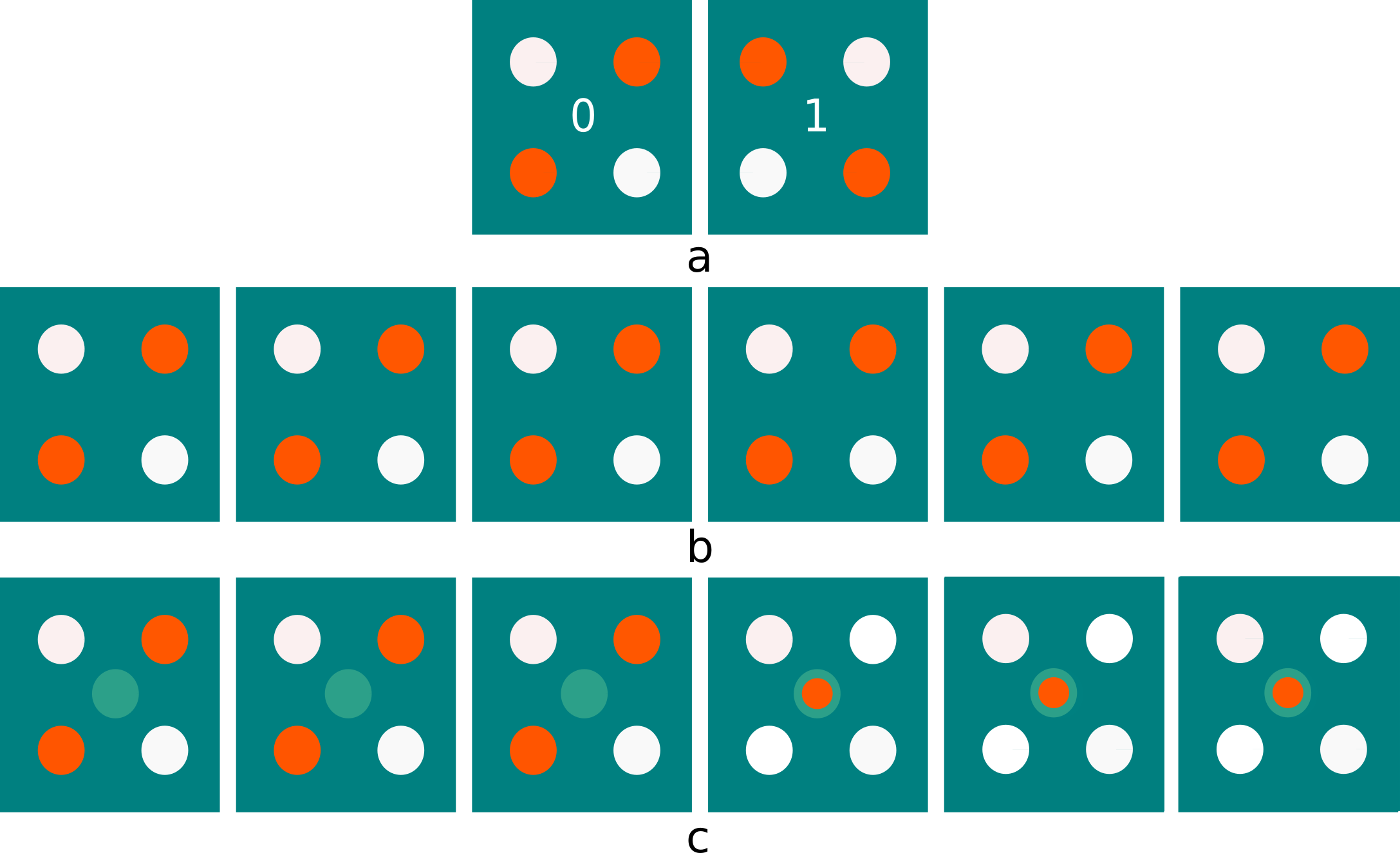}
  \caption{\footnotesize{\textit{a)} The two possible states of a typical QCA cell. 
  \textit{b)} A wire of simple QCA cells in one of its two  
  lower energy states. \textit{c)} A wire of 5-dots QCA cells, with its right side in the null-state, 
  as under the effect of a clocking 
  field. In this case it is possible to control the information flux direction.
  }}
  \label{fig:CGscheme}
\end{figure}

A number of theoretical investigations led to the recognition of some classes of promising candidate molecules for the 
implementation of molecular QCA\cite{lulent1,lulent2,boryl,lentisaksen}. Among these are a number of dinuclear 
mixed-valence metal complexes\cite{braun} mainly involving Fe and Ru, 
such as Creutz-Taube complexes\cite{tokunaga},
and various larger molecules containing tipically four metallic centers in a square, planar arrangement, forming a 
complete molecular QCA cell\cite{groizard,jiao}. Dinuclear species on the other hand constitute only half of a typical 
QCA cell, and can be arranged side by side on a surface in order to obtain a complete one\cite{italiani}.

\begin{figure}
  \includegraphics[width=0.44\textwidth]{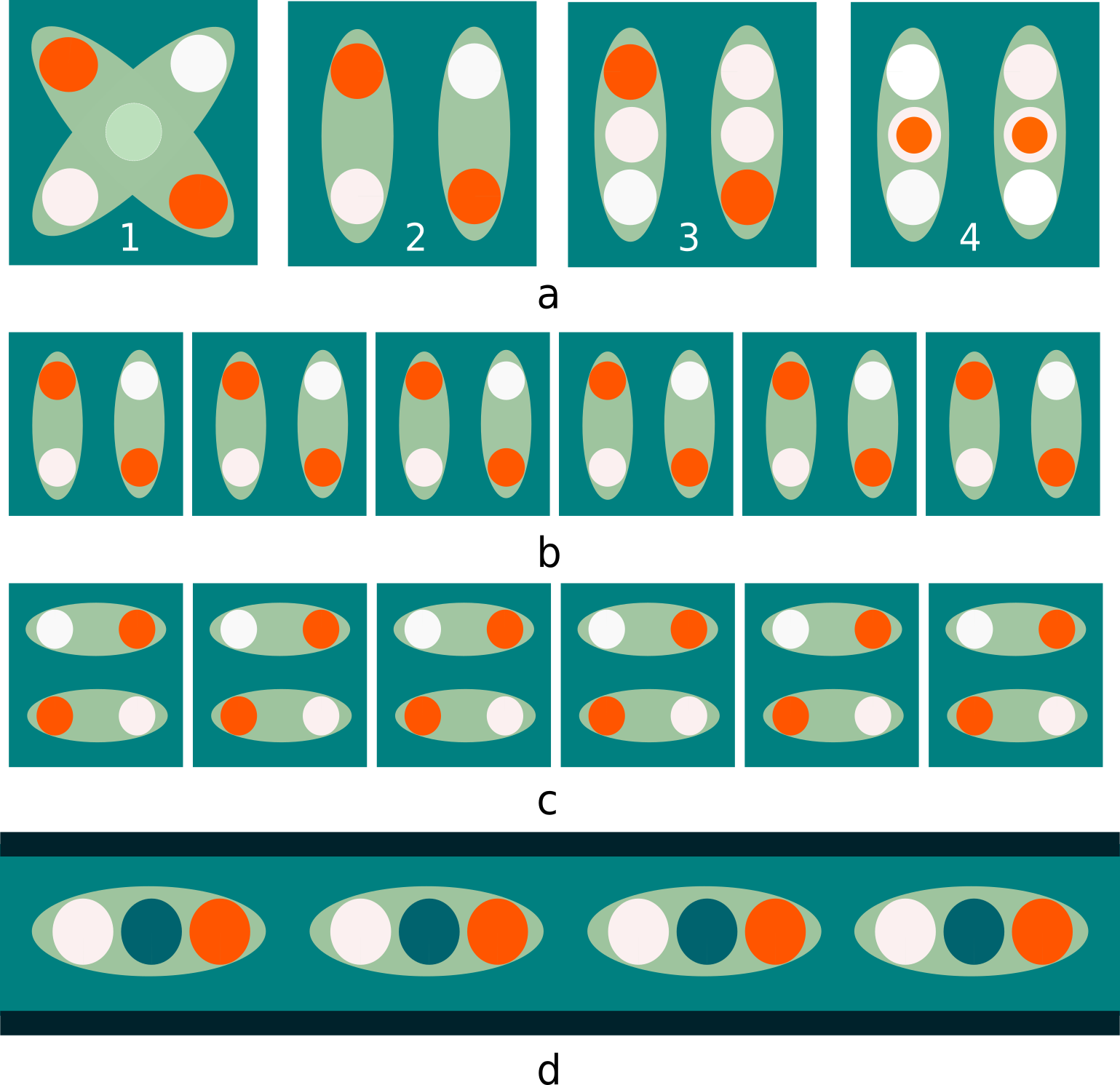}
  \caption{\footnotesize{
   Schemes of various kinds of molecular QCA cells and wires. \textit{a)} Complete cells as obtained from (1) 
  a single 4-dots molecule, possibly with a central site, whose occupation results in a null-state; (2) two paired 2-dots molecules;
  (3) two paired 3-dots molecules, resulting in a 6-dots cell, capable of assuming the null state (4).
  \textit{b)} A classical wire of molecular QCA cells with molecules assuming side-by-side orientation.
  \textit{c)} A wire consisting of head-to-tail oriented molecules, as proposed by Wang \textit{et al.}\cite{fullerene}
  \textit{d)} An encased wire consisting of a single row of head-to-tail oriented molecules as the one investigated in this paper.
  In all cases, white circles represent quantum dots and orange circles represent the moving charges, 
  while cyan circles in \textit{d)} 
  represent the countercharge for a zwitterionic molecule.
  }}
  \label{fig:CGscheme}
\end{figure}

The QCA machinery requires the metallic centers to be in different oxidation states with a 1:1 ratio, 
so that half of the metallorganic 
groups bear an extra charge, and 
the molecule (or pair of molecules) to be able to switch between two degenerate, or almost degenerate ground states as to 
minimize coulombic repulsion between charged groups. 
By immobilizing such molecules on a suitable surface in such a way that they form a pattern of dots like those shown in 
Fig. 2, and assuming 
that it is feasible to interact with them on the molecular scale for input output purposes, one can 
in principle create a computing QCA device, and modern techniques of electron beam lytography and tunnel electron microscopy seem to 
bring this goal at hand.\cite{}



The neutral, mixed valence diferrocenyl-carborane 7-Fc$^{+}$-8-Fc-7,8-\textit{nido}-[C$_{2}$B$_{9}$H$_{10}$], Fig. 3, on which we 
focus in this paper, was synthesized and characterized, along with some closely related species, 
by Christie \textit{et al.}\cite{lentmain}. They found convincing evidence of its zwitterionic nature, as the average 
Fe-C bond lengths, measured by X-ray diffraction, differ 
between the two ferrocenyl groups, suggesting that the carborane is able to oxidize one of the ferrocene groups to 
ferrocenium. The 
result of this \textit{self-doping}\cite{lulent} mechanism is a zwitterion with a 
formal +1 charge on one of the ferrocenyl groups and a -1 on the 
central carborane group, which could in principle act as an half six-dots QCA cell.

Dealing with a zwitterion is a great advantage, as counterions accompanying other switches proposed in literature, 
most of wich are charged radicals, must be taken care of both in calculations and in devising ipothetical applications.
A further advantage of this species lies in its excited, symmetric, non-zwitterionic form, resulting from the electron  
transfer between the negatively charged carborane cage and ferrocenium. 
Christie \textit{et al.} recognize this reversion of self-doping as responsible for a band around 11000 $cm^{-1}$ they observed 
in the UV-Vis spectrum of the molecule. As we show below the symmetric form, which can be termed the 
null-state for our species, is easily obtained by application of a suitable electric field, allowing in principle 
for efficient clocking.
\begin{figure}[h!]
  \includegraphics[width=0.30\textwidth]{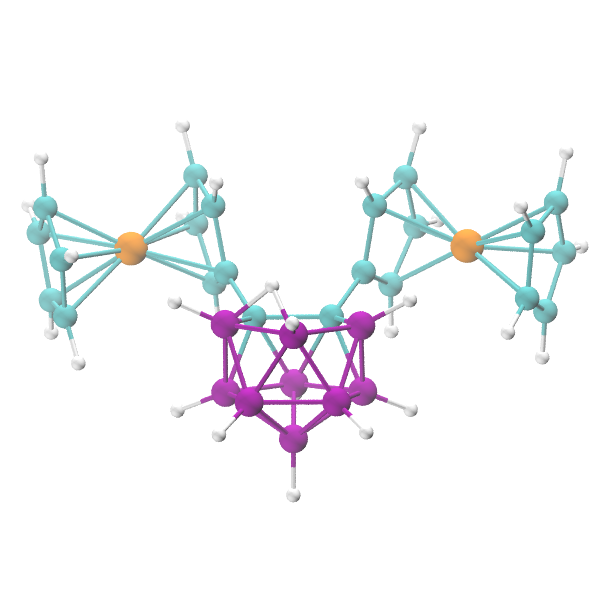}
  \caption{\footnotesize{7-Fc$^{+}$-8-Fc-7,8-\textit{nido}-[C$_{2}$B$_{9}$H$_{10}$], a recently synthesized diferrocenyl 
  carborane\cite{lentmain}. 
  Notice the asymmetric bridge H atom on carborane, defining what we call the $\alpha$-side of the molecule throughout this paper.
  (This and the following figures were generated using the VMD package\cite{vmd})
  }}
  \label{fig:CGscheme}
\end{figure}
The practical realization of all molecular QCA schemes proposed to date depend on the developement 
of suitable techniques 
capable of arranging and binding molecules on a surface according to the desired geometry. 
However, while various promising developement in this sense are reported in the literature, a truly effective technique is still missing
In this work we propose a different scheme resulting in the molecules assuming their functional position 
without external intervention, 
as they are adsorbed in the channels of zeolite ITQ-51\cite{martinez}. The framework of this microporous 
material shows large parallel channels, with an elliptical 
cross section of $\sim$ 13 x 9 \AA{}. This particular geometry fits quite tightly around diferrocenyl molecules, 
so that they are forced to 
move inside channels in a single row, resulting in what is termed single-file diffusion\cite{karger}, 
with their Fe-Fe axis roughly parallel to the channel 
axis, as depicted in fig. 4. 

Under this conditions they cannot overtake each other, nor fully rotate around any of their axis, and basically can only move 
back and forth 
between their first neighbours. In this way the system offers itself as a promising candidate for a QCA wire implementation: 
if, as our calculations demonstrate, the head-to-tail interaction between host diferrocenyl carborane molecules is capable of 
efficiently orienting the ground state of neighbours, one can easily envision a way of transmitting information along the loaded 
channel, 
resulting in a new kind of QCA wire, which consists of a single row of head-to-tail oriented 2-dots molecules. Such 
a wire model has not yet been studied, the only similar model being the one 
proposed by Wang \textit{et al.}\cite{fullerene,fullerene2}, with head-to-tail 
interactions in a double row of fluorinated fullerene, resulting in a wire of tilted 4-dots QCA cells.
While the classical 4-dots scheme allows for more complex operations as noted above, it has  
no particular advantage when dealing with the kind of wire proposed here.

\begin{figure}
  \includegraphics[width=0.34\textwidth]{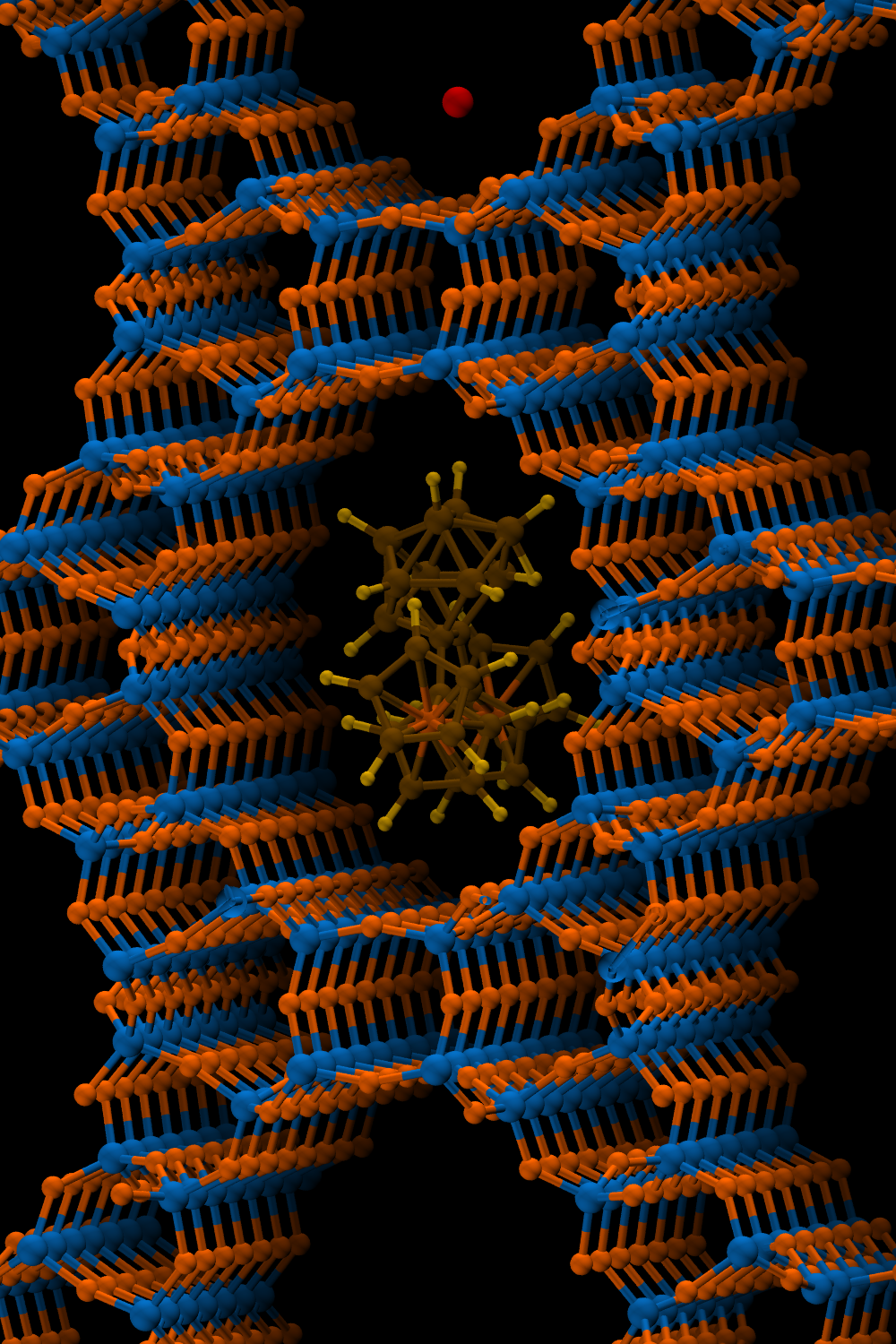}
  \caption{\footnotesize{A sketch of diferrocenyl-carborane inside one of the main channels of zeolite ITQ-51. The molecule can 
  only translate parallel to the channel axis, and full rotation around any axis is not possible. The external charge used to drive 
  the system to the null-state is also depicted in the upper channel (see text).
  }}
  \label{fig:CGscheme}
\end{figure}

There is however a point which deserves attention in devising a suitable clocking scheme for our system. With reference 
to Fig. 4, the clocking field should lie roughly along the vertical direction in order to push the excess electron on carborane 
towards the ferrocenium group.
However there are two 
symmetric orientation available to each molecule within the channels, with respect to this direction, and  
a simple sweep of the 
clock would not be effective unless one finds a way of orienting all molecules in the same way. 
If the molecule in 
fig. 4 had its carborane group pointing in the opposite direction, the negative charge would further stabilize a state of the molecule  
whith excess positive charge on its side, instead of triggering a switch to the null-state.  
So if molecules are randomly oriented with respect to the vertical direction,  some of them will retain 
their state, thus interfering with, and possibly stopping, any information propagation. 
Although such specific problems are not of the kind we are primarily concerned with in this paper, we point out that there should be 
ways of circumventing it: namely by orienting host molecules during adsorption with the application of a strong electric field, or by implementing 
more sofisticated clocking schemes wich alternates opposite field orientations. Artenatively one could also resort to other means 
of switching, not yet investigated to date, e.g. 
light absorption which, according to the aforementioned spectroscopic results of Christie \textit{et al.}\cite{lentmain}, 
is able to transiently excite the diferrocenyl-carborane to the null-state.


 
\section{DFT calculations and results}

We investigated the electronic structure of diferrocenyl-carborane by means of Density Functional Theory  
in the Kohn-Sham formulation, 
using the ORCA free package\cite{orca}. 
Geometry and orbital optimizations were performed with the test molecule in various environments, in order to test its capabilities as 
a molecular switch, the nature of the ground state in vacuum and within the confining environment, and the conditions, if any, 
under which it switches to the null-state. In all cases reported the B3LYP\cite{becke} functional was used, 
with the def2-TZVP basis set\cite{def2}, and the D3 
correction for dispersion interactions. 
Polarization functions were used in preliminary calculations, resulting in 
a drastic increase of computational time, while only negligibly affecting the properties of interest, so we did not use 
them in production runs. As the molecule investigated is a radical, all calculations were conducted within the open-shell 
unrestricted Kohn-Sham scheme.
\begin{table}[!h]
  \begin{tabular}{|c|c|c|c|c|}
    \hline
      & Fe-C($\alpha$) & Fe-C($\beta$) & Fe-C($\alpha$)* & Fe-C($\beta$)*  \\
    \hline
    Single & 2.1392 & 2.0832 & 2.1386 & 2.0829 \\
    \hline
    Pair($\alpha$) & 2.0878 & 2.1327 & 2.0804 & 2.1297 \\
    \hline
    Pair($\beta$) & 2.1395 & 2.0831 & 2.1388 & 2.0828 \\
    \hline    
    Null & 2.0870 & 2.08487 & 2.0862 & 2.0839 \\
    \hline 
  \end{tabular}
  
\caption{Fe-C bond lenghts for the two ferrocenyl groups, in \AA{}. 
The ($\alpha$) refers to the Fc group or driver being on the side of 
the asymmetric H atom of carborane, and ($\beta$) refers to the opposite side. Data in starred columns refer to the 
nanoconfined setting.}
  \label{tab1}
\end{table}

Quantum calculations under nanoconfinement were performed on the single molecule, with the environment 
of the ITQ-51 channel, within a radius of 21 \AA{}, simulated by a collection of fixed point charges, 
namely -0.5 a.u. for O and 1.0 a.u for Si atoms, 
as obtained in a previous work on ITQ-29, a similar aluminosilicate\cite{gabrieli}. 
The neighboring 
molecule, acting as driver, was represented by the Loewdin charges 
previously obtained for the single molecule in vacuum, the distance between the centers of mass of the the two molecules being  
13 \AA{} in all cases reported. The driver molecule was located on one side of the reference one as depicted in Fig. 5 with 
the ferrocenium group pointing towards the test molecule. 
Calculations on all arrangements of the test molecule and driver charges were performed both   
under confinement and in vacuum, by adding or removing, respectively, the framework of Si and O charges only.
\begin{table}[!h]
  \begin{tabular}{|c|c|c|c|c|c|c|}
    \hline
      & Fc($\alpha$) & Fc($\beta$) & C$_{2}$B$_{9}$H$_{10}$ & Fc($\alpha$)* & Fc($\beta$)* & C$_{2}$B$_{9}$H$_{10}$*  \\
    \hline
    Single & 0.850 & 0.0733 & -0.923 & 0.863 & 0.085 & -0.949 \\
    \hline
    Pair($\alpha$) & 0.054 & 0.868 & -0.923 & 0.061 & 0.886 & -0.947 \\
    \hline
    Pair($\beta$) & 0.860 & 0.056 & -0.917 & 0.870 & 0.069 & -0.939 \\
    \hline    
    Null & 0.002 & 0.028 & -0.030 & 0.010 & 0.040 & -0.050 \\
    \hline
  \end{tabular}
  
\caption{Loewdin cumulative charges on Fc and carborane groups, in a.u. 
The ($\alpha$) refers to the Fc group or driver being on the side of 
the asymmetric H atom of carborane, and ($\beta$) refers to the opposite side. Data in starred columns refer to the 
nanoconfined setting.}
  \label{tab1}
\end{table}
Our results confirm that the ground state for the single molecule in vacuum, as well as 
inside the ITQ-51 channel, is the zwitterionic form, always showing the ferrocenium group on the same 
side of the asymmetric bridge H atom on carborane, to which we will refer as $\alpha$-side.

The presence of the second molecule lead to the reorientation of the dipole moment both in vacuum 
and under confinement, as to maximize the distance between the two positively charged ferrocenium groups, 
while no switch is observed when the 
driver molecule has its ferrocenium group pointing away from the test. 
\begin{figure}
  \includegraphics[width=0.44\textwidth]{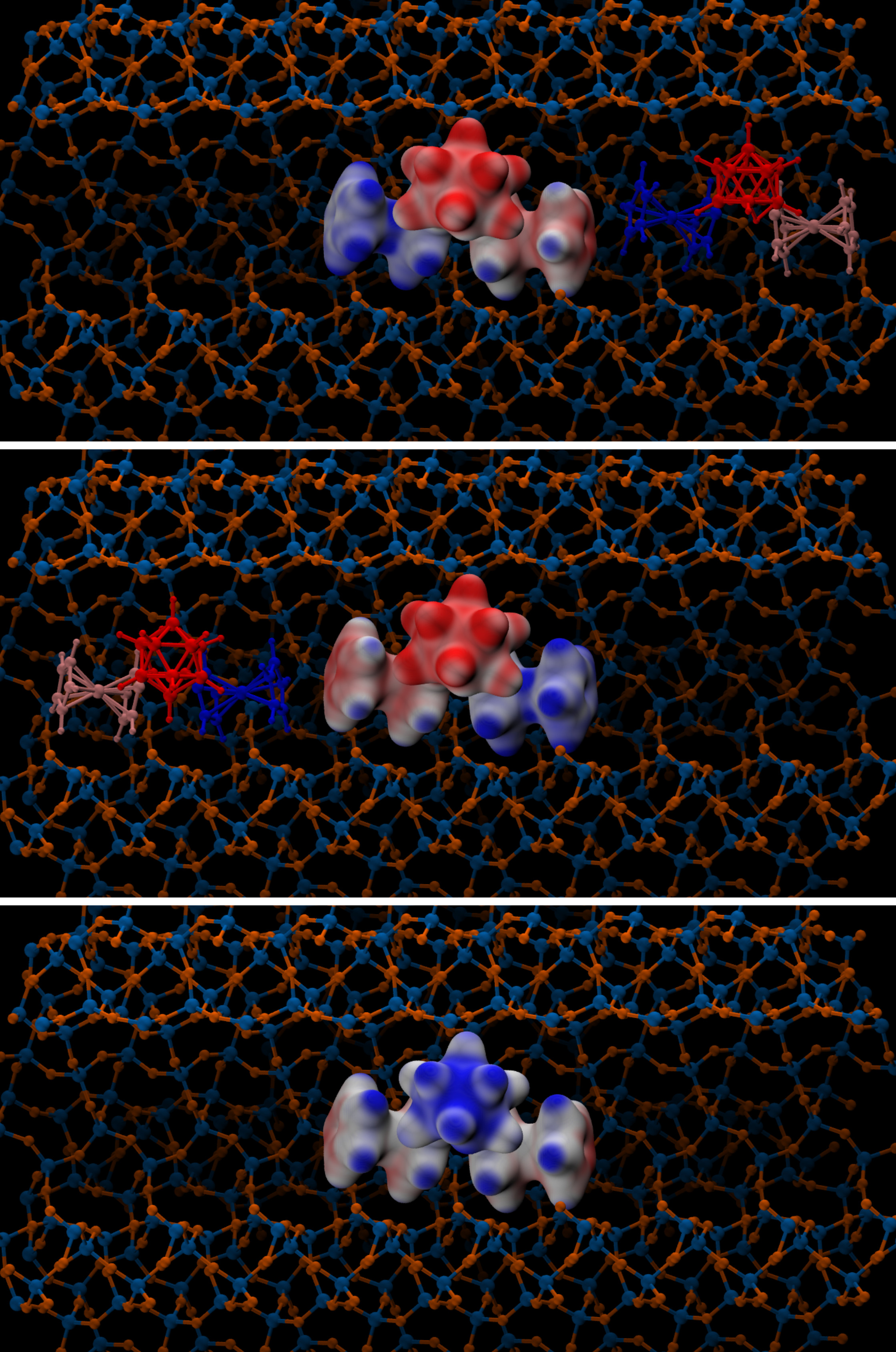}
  \caption{\footnotesize{Electrostatic potential map for diferrocenyl carborane under various external charge settings.
  The top figure depicts the Pair($\alpha$) configuration, with the driver molecule on the same side, ($\alpha$), of the 
  asymmetric H atom on the test molecule, while the figure in the middle shows the specular, ($\beta$), situation. In 
  both cases the color scheme for the driver refers to the overall rough charge of the three groups: blue and red stand for positive 
  and negative charge respectively, and pink for neutral. The bottom figure shows the null-state, as obtained from the charge 
  configuration in Fig. 3. Part of the zeolite ITQ-51 framework, as well as the negative driver charge in 
  the bottom figure were omitted for clarity.  
  }}
  \label{fig:CGscheme}
\end{figure} 
The molecule switch is also confirmed by total Loewdin charges on carborane and on the 
two ferrocenyl groups, as well as by the average Fe-C bond lenghts 
in the latters, as reported in Tab. I. These are consistent with our DFT results 
for isolated ferrocene and ferrocenium, 2.084 and 2.134 \AA{} respectively, and slightly longer than the experimental values, 2.05 
and 2.095 respectively,  
reported by Lent \textit{et al.}\cite{lentmain}, while the relative difference is almost the same. 
Another clear signal of the asymmetry between the two groups is given by the free valence, 
as obtained from Mayer population analysis\cite{mayer}, for the Fe atoms. This is invariably of $\sim 0.9$ for Fe(III), 
and $\sim 0$ for Fe(II), again in agreement with our DFT results for isolated ferrocene and ferrocenium.

The null-state of diferrocenyl-carborane was also obtained, by using a single external charge 
varying in the range from -10 to -3 $e$, located at $\sim$ 14 \AA{} 
from the center of mass of diferrocene, 
as illustrated in Fig. 4. This arrangement mimics the effect of a clocking field 
perpendicular to the channel axis, and results in an electron transfer from carborane to ferrocenium. 
The transition is made evident by a strong decrease in the magnitude of the dipole moment, and of the cumulative Loewdin 
charges of the three groups in the molecule, all of which fall to nearly zero. Moreover the Fe-C average bond lengths in the 
two groups becomes equal, and both Fe atoms assume a null free valence.  
\begin{table}[!h]
  \begin{tabular}{|c|c|c|c|c|}
    \hline
      & Fe($\alpha$) & Fe($\beta$) & Fe($\alpha$)* & Fe($\beta$)* \\
    \hline
    Single & 0.9355 & 0 & 0.9407 & 0 \\
    \hline
    Pair($\alpha$) & 0 & 0.9457 & 0 & 0.9533 \\
    \hline
    Pair($\beta$) & 0.9395 & 0 & 0.9434 & 0 \\
    \hline    
    Null & 0 & 0 & 0 & 0 \\
    \hline
  \end{tabular}
\caption{Free valences according to Mayer for Fe atoms. 
The ($\alpha$) refers to the Fe atom or driver being on the side of 
the asymmetric H atom of carborane, and ($\beta$) refers to the opposite side. Data in starred columns refer to the 
nanoconfined setting.}
  \label{tab1}
\end{table}
A comparison of the results for the molecule under confinement and in vacuum, and 
inspection of the energy values, shows that 
the zwitterionic form is somehow stabilized under confinement and its charge separation 
slightly more pronounced. By contrast the effect on the null-state is an almost negligible 
destabilization and blurring of charge distribution. This implies that a stronger field could 
be needed in order to clock the system within the zeolite channels.





\begin{table}[!h]
  \begin{tabular}{|c|c|c|c|c|}
    \hline
      & $\mu$ & $\mu$* & E & E* \\
    \hline
    Single & 17.53 & 17.95 & -3605.8544 & -3605.8677 \\
    \hline
    Pair($\alpha$) & 17.41 & 18.80 & -3605.8604 & -3605.8727 \\
    \hline
    Pair($\beta$) & 18.29 & 18.58 & -3605.8569 & -3605.8692\\
    \hline    
    Null & 11.70 & 11.35 & -3605.8819 & -3605.8803 \\
    \hline
  \end{tabular}
  
\caption{Modules of the dipole moment $\mu$ (in Debye), and energies (in Hartree). 
Data in starred columns refer to the 
nanoconfined setting.}
  \label{tab1}
\end{table}



\section{Conclusions \label{sec:conclusions}}
 
We reported DFT calculations results for a diferrocenyl carborane which was recently proposed as a promising molecular switch, 
and found it to be suitable 
for use in QCA circuitry at the nanometric scale. 
Our theoretical findings are in agreement with the spectroscopic and crystallographic results previously reported in literature, and 
support the switchable nature of this species, not only in vacuum or solution, but also in a nanoconfining environment, namely within 
the channels of zeolite ITQ-51. 

The vast class of microporous materials is constantly growing, and synthetic techniques are getting more and more sofisticated, 
attaining tailored geometries for disparate purposes, while ship-in-a-bottle methods allow complex molecules 
to be formed inside pores from smaller adsorbed precursors\cite{shipinbottle}. 
So if these materials can host molecular switches and act 
as an embedding matrix for QCA architectures, without hindering their bistability, switchability, and null-state properties, 
a wealth of interesting opportunities becomes available. 
We investigated only QCA wires within a framework of straight channels, as this is the simpler and more easily handled system for 
a preliminary study, as well as for future experimental tests. However nothing prevents in principle future developments 
towards more complex schemes, perhaps taking advantage of the three dimensional rich structure of microporous materials, which, 
while posing challenging problems, could result in an exciting enrichment of QCA possible range of implementations. 
Keeping molecular switches within the channels 
once they have been adsorbed, and interacting with them for input-output operations, constitute crucial problems in view 
of any practical application of a zeolite embedded QCA. 
In this sense we find inspiring the work of Calzaferri and coworkers\cite{calzaferri0,calzaferri,calzaferri2,calzaferrimain}, 
who proposed a number of stop-cock molecules capable of 
binding to the entrance of the zeolite pores and lock inside adsorbate molecules. 
They could also be very useful in input-output operations, if one succedes in making a stop-cock which is a molecular 
switch itself. Moreover as these authors show, it is also possible to link various crystals together, which could be 
a viable way of assembling longer wires, and ensuring communication between them, 
as they demonstrate it is possible for electronic excitations.
Clearly many aspects still need to be investigated, and in particular the dynamics of the information transfer, 
both at the level of intramolecular charge migration, and of intermolecular interactions. In this sense, 
it would be desirable to develope a proper force field for sensible classical Molecular Dynamics simulations, 
in order to investigate single file diffusion within zeolite channels, and the 
frequency of encounters between molecules close enough to allow informations transfer. On the other hand more sophisticated 
methods, such as  QMMM calculations, would probably be needed for a quantitative understanding of the switching 
kinetics\cite{blairdynam,luliu,tougaw} in such a complex supramolecular environment.




\bibliographystyle{aipnum4-1}
\bibliography{bibliografia2.7}

\end{document}